\documentclass[prb,twocolumn,showpacs,preprintnumbers,amsmath,amssymb,floats,citeautoscript,nobalancelastpage]{revtex4-1}
\usepackage{graphicx}
\usepackage{dcolumn}
\usepackage{bm}
\usepackage{txfonts}

\begin{document}
 
\title{Current-Induced Gap Suppression in the Mott Insulator Ca$_2$RuO$_4$}

\author{R.~Okazaki$^{1,\ast}$}
\author{Y.~Nishina$^{1}$}
\author{Y.~Yasui$^{1,\dag}$}
\author{F.~Nakamura$^{2}$}
\author{T.~Suzuki$^{2}$}
\author{I.~Terasaki$^{1}$}

\affiliation{$^1$Department of Physics, Nagoya University, Nagoya 464-8602, Japan}
\affiliation{$^2$ADSM, Hiroshima University, Higashi-Hiroshima 739-8530, Japan}

\begin{abstract}
We present nonlinear conduction phenomena in the Mott insulator Ca$_2$RuO$_4$
investigated with a proper evaluation of self-heating effects. 
By utilizing a non-contact infrared thermometer, the sample temperature was accurately 
determined even in the presence of large Joule heating. 
We find that the resistivity continuously decreases with currents under an isothermal environment.
The nonlinearity and the resulting negative differential resistance occurs at relatively low current range,
incompatible with conventional mechanisms such as hot electron or impact ionization.
We propose a possible current-induced gap suppression scenario, 
which is also discussed in non-equilibrium superconducting state or charge-ordered insulator.
\end{abstract}

\maketitle

Nonlinear transport nature of strongly correlated electrons is one of the most fundamental 
but remaining issues in condensed matter physics.
In a vicinity of correlated insulating phase, mobile electrons sense strong interactions among them 
and consequently anomalous metallic states are often realized,
which are usually induced by temperature change, physical pressure or chemical substitutions \cite{Imada98}.
This naturally invokes an idea that 
the correlated electrons in a highly non-equilibrium condition,
such as in strong electric field,
show exotic behaviors as well \cite{Oka03}.
In correlated transition-metal oxides or organic salts,
such a nonlinear conduction phenomenon has been extensively explored  \cite{Potember79,Tokura88,Asamitsu97,Taguchi00,Zeng08}.
In an oxide Mott insulator, temperature variation of the  threshold field for dielectric breakdown is found to be similar to that in the charge-density-wave (CDW) materials,
implying a possible collective motion triggered by strong fields \cite{Taguchi00}.
Indeed, a spontaneous electrical oscillation associated with notable nonlinear conduction has been 
reported in an organic charge-order salt \cite{Tamura10},
which is reminiscent of the sliding motion of CDW. 
As a different origin for the breakdown phenomena, 
an unconventional avalanche process with anomalously long delay time has been suggested 
in the narrow-gap chalcogenide Mott insulators \cite{Guiot13}. 

The 4$d$-electron Mott insulator Ca$_{2}$RuO$_4$ \cite{Nakatsuji97,Cao97} 
is a particularly suitable example 
for the study of nonlinear transport nature in correlated electron systems 
because the insulating phase of this material is highly susceptible to external perturbations such as 
heating, application of pressure, or chemical substitution \cite{Alexander99,Nakatsuji00L,Nakatsuji00,Nakamura02}.
At $T_{\rm MI} \simeq$ 360~K, 
this compound exhibits a first-order metal-insulator transition,  
whose nature has been intensively studied as orbital order formation \cite{Mizokawa01,Hotta01,Anisimov02,Lee02,Jung03,Kubota05,Zegkinoglou05,Gorelov10}.
Systematic isovalent Sr substitution study has revealed that 
the ground state of Ca$_{2-x}$Sr$_{x}$RuO$_4$ varies from 
the Mott insulator ($x<0.2$) to the spin-triplet superconductor ($x$ = 2) \cite{Maeno94}
through a spin-glass state in the broad composition range \cite{Carlo12}.
The parent compound becomes to be metallic with applying pressure as well \cite{Nakamura02}, and 
the higher pressure makes the system superconducting \cite{Alireza10}.

Recently, Nakamura {\it et al.} reported an electric-field-induced insulator-to-metal transition in Ca$_{2}$RuO$_4$ \cite{Nakamura11}.
They have found that the Mott insulating state is abruptly switched into a metallic one by relatively low electric field of $E \sim$ 40 V/cm at room temperature, 
indicative of a novel mechanism for this transition.
They used, however, a relatively long pulse with duration time of 10 to 100 ms, 
raising a serious concern that the observed field-induced switching is just from the Joule heating effect
since $T_{\rm MI}$ is close to room temperature.
The duration time is indeed much longer than that in
traditional nonlinear-conduction experiments with a short pulse less than 1 $\mu$sec \cite{Ryder53,Glicksman58,Bauer72}.
The heating effect is the most fundamental problem in such experiments
because the conductive state realized by strong field is often very similar to that in high temperatures.
In fact, the origins of observed nonlinear conduction in correlated electron systems have been 
discussed as the heating effect \cite{Sacanell05,Song06,Carneiro06,Hucker07,Kim10,Zhong11,Zimmers13}.

Here, we investigate the nonlinear transport property of the Mott insulating phase of Ca$_{2}$RuO$_4$
using a proper temperature evaluation method.
This is achieved with utilizing infrared thermometer, 
which directly measures the sample temperature without contacts, 
in contrast to conventional contact thermometry by which we cannot avoid a finite temperature difference between the sample and thermometer
due to the thermal resistance.
Using the non-contact technique, the sample temperature was strictly kept below $T_{\rm MI}$ even in the large external currents.
We find a continuous reduction of the electrical resistivity by the current
and a distinct negative differential resistance behavior at very low fields, indicating that
the conventional mechanisms for dielectric breakdown in semiconductors are unlikely in this system.
We propose a model that the energy gap is suppressed by external current 
as suggested in non-equilibrium superconducting state or charge-ordered insulating phase.
A microscopic mechanism for the current-induced gap suppression is discussed
in terms of orbital order melting by external current.

\begin{figure}[t]
\includegraphics[width=1\linewidth]{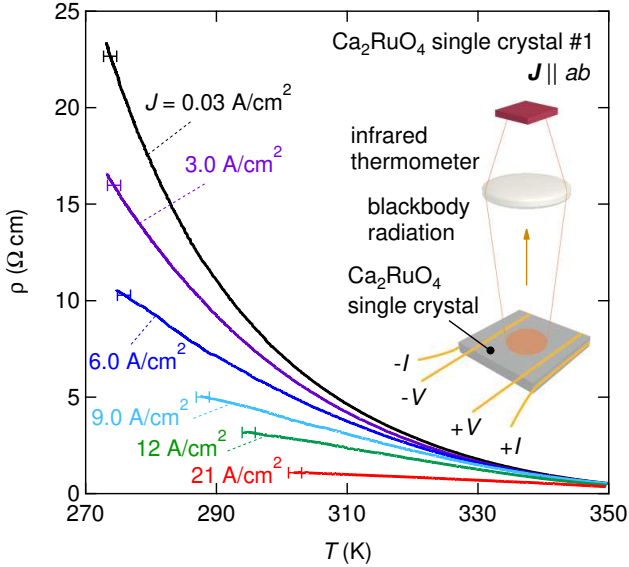}
\caption{(Color online).
Temperature dependence of the in-plane resistivity $\rho$
in the Mott insulating phase of Ca$_2$RuO$_4$ measured with different external current densities $J$. 
The sample temperature (horizontal axis) was determined by a non-contact infrared thermometer. 
The error bars  show the temperature measurement resolution of the infrared thermometer.
Inset illustrates the schematic figure of experimental configuration. 
The infrared thermometer locates just above the single-crystalline sample.}
\end{figure}

An important issue on the nonlinear transport is a reliable determination of the sample temperature 
even in large external currents.
In the present study, to evaluate the sample temperature with heating,  
we utilized a non-contact infrared thermometer (Infrared temperature sensor IT2-02, Keyence Corporation),
which measures the temperature through the 
blackbody radiation from the sample in an infrared range ($\lambda = 6 \sim$ 12 $\mu$m).
This method enables us to directly measure the temperature without additional heat capacity and contact thermal resistance
of conventional contact-type thermometer. 
In our experiments, Ca$_2$RuO$_4$ single-crystalline sample was suspended using four gold wires and 
thus isolated from the thermal bath to avoid a strong temperature gradient inside the sample.
The experimental setup was built in a N$_2$-gas-filled glove box to prevent the sample from icing and
the external temperature was controlled by N$_2$ gas blowing.
Just above the sample, the infrared thermometer was placed as illustrated in the inset of Fig.~1.
The spot-size diameter of the infrared thermometer is 1.2 mm, 
which is smaller than the cleaved $ab$-plane surface area (typical sample dimensions of $2\times2\times0.2$\,mm$^3$).
For the calibration of the emissivity to use the thermometer, 
the sample surface was thinly coated by a blackbody insulating paste.
The accuracy of infrared thermometer was checked using a calibrated thermometer coated by the paste.
The temperature values measured by the two methods are well accorded within 1 K.
This also indicates a negligible temperature difference between the sample and the paste.
The temperature homogeneity in large currents was checked by an infrared thermography (InfReC R300, Nippon Avionics).
The variation is about 1 K, which is almost the same as the measurement resolution of the spot-type infrared thermometer.
The in-plane resistivity was measured as a function of temperature using the standard four-probe dc method
with several different values of external constant currents for {\boldmath $J$}$\parallel ab$-planes. 
To reduce a contact resistance that possibly becomes an extrinsic origin for nonlinear conduction due to local heating, 
electrical contacts were carefully made with a gold deposition technique. 
Ca$_2$RuO$_4$ single crystals with essentially stoichiometric oxygen content were grown with a flouting-zone method. 

Figure\:1 depicts the temperature dependence of in-plane resistivity $\rho$ in Ca$_2$RuO$_4$ single crystal 
measured with several excitation currents.
Note that the temperature was directly determined by the non-contact infrared thermometer,
whose temperature measurement range is between 273 and 773 K.
The high-current resistivity data is not available at lower temperatures since the sample was not cooled owing to large heating.
At  $T_{\rm MI}$, it is highly probable that the crystal will be broken owing to the large structural change \cite{Alexander99}, and
thus we kept the sample temperature below $T_{\rm MI}$ during the experiment.
At $J=0.03$ A/cm$^2$, the resistivity behavior is the same as that in the previous reports.
The resistivity measured with lower current density than 0.03 A/cm$^2$ is almost unchanged.
At higher external currents, while all the data share an insulating feature, namely $d\rho/dT<0$, in common, 
the resistivity clearly decreases with increasing external currents, 
indicating a nonlinear conduction from insulating toward a metallic state. 
The resolution of the infrared thermometer is shown as the error bars for each data.
Clearly, the observed nonlinearity is much larger than magnitudes of the error bar.

\begin{figure}[t]
\includegraphics[width=1\linewidth]{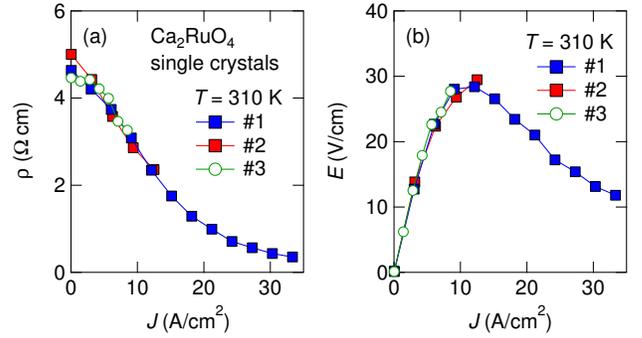}
\caption{(Color online).
Current density $J$ variations of (a) the resistivity $\rho$ and (b) the electric field $E$ at $T=310$ K. 
Results of three single crystals are shown as different symbols.
}
\end{figure}

In Figs.~2(a) and 2(b), we summarize the transport properties of Ca$_2$RuO$_4$ at 310 K 
as a function of the electrical current density extracted from Fig.~1.
Results of three Ca$_2$RuO$_4$ samples are shown with different symbols.
Essentially same results among these samples indicate no filamentary effect.
As seen in Fig.~2(a), the resistivity continuously decreases with currents,
and at the highest current value that we can achieve, 
the resistivity becomes about ten times smaller than the lowest current one.
This strong suppression of the resistivity leads to a negative differential resistance 
in the electric field $E$ vs. $J$ curves shown in Fig.~2(b).
At 310 K, the $E$-$J$ curve exhibits a peak at $E\sim30$ V/cm,
which is almost same as the reported threshold field \cite{Nakamura11},
indicating an intrinsic nonlinearity in this material.
Note that
the resistivity shows no abrupt change above this field 
but rather continuously decreases, still keeping an insulating behavior as seen in Fig.~1.
In this situation, the control parameter is essential:
we controlled the current instead of the electric field to avoid the thermal runaway.
Furthermore, the electric field is now a single-valued function of current while the current is a multiple-valued one of field,
implying that the current is more appropriate control parameter.

Let us discuss an origin of the observed nonlinear conduction phenomena in Ca$_2$RuO$_4$.
In the Zener tunneling mechanism for dielectric breakdown, the threshold field is roughly estimated as 
$\sim \varepsilon_g/ed$, where $e$ is the charge of an electron, $\varepsilon_g$ is a gap energy, and $d$ is a distance between neighboring atoms.
In Ca$_2$RuO$_4$, using $\varepsilon_g \sim 0.4$~eV \cite{Puchkov98,Nakatsuji04} and
$d \sim 0.4$~nm \cite{Braden98,Fridet01}, the threshold field becomes about 10 MV/cm, 
which is much larger than the applied field in the present study.
The many-body effect has an influence on the tunneling as the threshold is modified as
$\sim \Delta_{\rm Mott}/2e\xi$,
where $\Delta_{\rm Mott}$ is a Mott gap energy and $\xi$ is a doublon-hole correlation length \cite{Oka12},
but this still seems to be larger than the present field range.
In analogy to the sliding motion of CDW, 
a collective trasnport has been suggested in a Mott insulator \cite{Taguchi00}.
In such a case, the sliding CDW carries excess current above the threshold field.
However, the present system shows rather smooth change of the resistivity as a function of current,
implying a different mechanism.

In conventional semiconductors,
nonlinear transport is generally observed even in much lower field than the above-mentioned threshold field.
This is so-called hot electron phenomenon, in which 
the electronic temperature $T_e$, introduced from the kinetic energy of electron $\varepsilon_{\rm kin}$ as $T_e\equiv\varepsilon_{\rm kin}/k_B$,
is larger than the lattice temperature $T_l$ \cite{Conwellreview}.
In the acoustic phonon scattering regime with the energy-dependent relaxation time of $\tau(\varepsilon)\propto \varepsilon^{-1/2}$,
an increase of $T_e$ gives a decrease of $\tau$,  
thus the conductivity or the mobility $\mu$ is decreased with increasing field as observed in Si or Ge samples \cite{Ryder53}.
This picture is incompatible with the present result since the conductivity is increased with current as seen in 
Fig.~2(a).
In impurity-doped semiconductors, on the other hand, 
the electric field raises the mobility owing to the ionized impurity scattering with 
energy-dependent relaxation time of $\tau(\varepsilon)\propto \varepsilon^{3/2}$  \cite{Conwellreview}.
But even in this case, the nonlinearity is usually too weak to exhibit a negative differential resistance and
the impurity scattering is generally dominant only at low temperatures.

We next discuss the impact ionization mechanism.
To cause this,
it is necessary that the electron kinetic energy $\varepsilon_{\rm kin}$ exceeds the gap energy $\varepsilon_g$, 
leading to a condition that the electron velocity $v=J/ne$ should be comparably large to 
$v_m = \sqrt{2\varepsilon_{g}/m^*}$, where $m^*$ is an effective mass \cite{Wolff54}.
In Ca$_2$RuO$_4$, assuming $m^*=m_0$, one obtains $v_m = 3.7\times10^7$ cm/s.
On the other hand, at 310 K, the carrier concentration is estimated to be $n=3.5\times10^{16}$ cm$^{-3}$, 
which is calculated from the measured Seebeck coefficient (not shown), 
leading to $v = 5.3\times 10^3$ cm/s at $J=30$ A/cm$^2$.
Now, $v \ll v_m$, indicating that the impact ionization mechanism is unlikely in the present system.
This situation should be compared with that in the conventional narrow-gap semiconductor InSb, 
in which a comparable energy gap of 0.2 eV is open.
In pure InSb samples, 
an avalanche breakdown  
occurs at relatively low fields of the order of 100 V/cm at 77 K
through the impact ionization associated with a drastic increase of the carrier concentration \cite{Glicksman58,Bauer72}.
This is due to a very high mobility of this material about 10$^5$ cm$^2$/Vs at 77 K, 
which is much larger than that in Ca$_2$RuO$_4$.
Also, the present field range is much smaller than the avalanche breakdown fields 
in the comparable narrow-gap Mott insulators \cite{Guiot13}.

The  continuous decrease of the resistivity observed at very low current (or field) range in Ca$_2$RuO$_4$ 
is highly anomalous compared with above-mentioned conventional mechanisms.
Here we analyze it in terms of a current-dependent energy gap, 
which has been proposed in non-equilibrium superconducting state \cite{Owen72}.
The superconducting gap is decreased not only by heating but also with excess quasiparticles injected by external currents,
as demonstrated by the thin film tunneling experiments \cite{Fuchs77,Akoh82}.
This carrier injection by current to reduce the energy gap is also discussed in CDW and charge-order insulators \cite{Sawano05,Ajisaka09}.
This scenario seems to be applied in the present case:
In Ca$_2$RuO$_4$,  
the tetragonal crystal-field splitting due to the flattened RuO$_6$ octahedra, 
as well as the strong Coulomb repulsion among $4d$ electrons,
leads to a stabilized $d_{xy}$ orbital ($d_{xy}$ orbital ordering) below $T_{\rm MI}$ \cite{Jung03}.
This is 
characterized by the orbital polarization $p\equiv n_{xy}-(n_{yz}+n_{zx})/2$, where  $n_{i}$ ($i=xy,yz,zx$) is the $i$-orbital occupation.
Below $T_{\rm MI}$, $p\sim1$ [$(n_{xy},n_{yz},n_{zx})=(2,1,1)$] is realized.
Here the excess carriers injected by external currents should decrease the orbital polarization. 
In consequence, the energy gap is reduced by current through the Jahn-Teller distortion of RuO$_6$ octahedra.

\begin{figure}[t]
\includegraphics[width=1\linewidth]{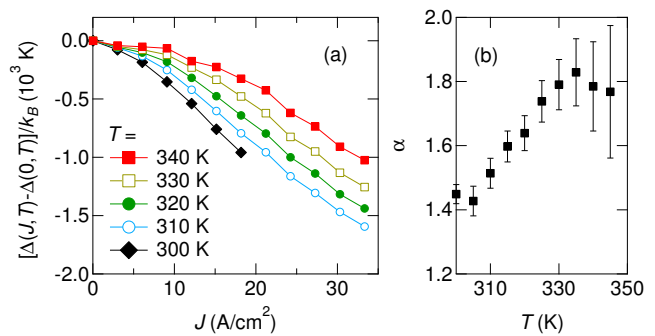}
\caption{(Color online).
(a) The energy gap change $[\Delta(J,T)-\Delta(0,T)]/k_B$ at various temperatures as a function of current density $J$ in sample \#1.
(b) Temperature dependence of the exponent $\alpha$, 
obtained from the fitting of the current-dependent gap using an expression  $\Delta(J,T)-\Delta(0,T)=AJ^{\alpha}$.
}
\end{figure}

In Ca$_2$RuO$_4$, the resistivity shows a slight deviation from activation-type insulating behavior near $T_{\rm MI}$,
whereas it is well obeyed below 250 K \cite{Nakatsuji04}, 
probably because of the temperature variations of the effective mass and the relaxation time.
Then the resistivity can be represented as
\begin{equation}
\rho(J,T) = \rho_0(T)\exp[\Delta(J,T)/2k_BT]
\end{equation}
where $\rho_0(T)$ is a temperature-dependent coefficient and $\Delta(J,T)$ is the current-dependent energy gap.
The gap also depends on temperature linearly near 300 K \cite{Jung03}.
Using Eq.~(1), one obtains the energy gap change from the low-current value as
\begin{equation}
\Delta(J,T)-\Delta(0,T) =2k_BT\ln \frac{\rho(J,T)}{\rho(0,T)}.
\end{equation}
Figure~3(a) depicts the current density dependence of the energy gap change
calculated using Eq.~(2) for various temperatures.
The gap shows a power-law variation expressed as $\Delta(J,T)-\Delta(0,T)=AJ^{\alpha}$. 
Here $A$ is a constant and $\alpha$ is an exponent, whose temperature variation is shown in Fig.~3(b).
In non-equilibrium superconducting or charge-order state, the normalized energy gap $\Delta(n_{\rm ex})/\Delta(0)$ is decreased 
as $\Delta(n_{\rm ex})/\Delta(0) = 1-2n_{\rm ex}$ for $n_{\rm ex}\ll1$ at low temperatures,
where $n_{\rm ex}$ is the excess quasiparticle density normalized by the condensed carrier density \cite{Owen72,Ajisaka09}.
Here $n_{\rm ex}$ is almost proportional to $J$, thus the linear reduction of gap as a function of current is expected.
As seen in Fig.~3(a), our present results show a similar behavior at lower temperatures, 
supporting the scenario that the gap is suppressed by current.
Close to the transition temperature, on the other hand, $J^2$-dependent gap reduction is 
theoretically found at low current range \cite{Ajisaka09}, also consistent with our result.
The gap is less reduced by current in higher temperatures since
the current-induced carrier injection effect is relatively weakened in the high-temperature range where a large number of carriers are thermally excited.

A remaining question is that why the nonlinear conduction appears in such a low field range.
A characteristic length for the nonlinearity is estimated as $l = \delta\Delta/eE_0 \simeq 10$ $\mu$m,
where $\delta\Delta \simeq 30$ meV is a gap reduction at the current $J_0 = 10$ A/cm$^2$ and the field $E_0\simeq 30$ V/cm [Fig.~3(b)].
In Ca$_2$RuO$_4$, the electron-phonon coupling is significantly large, 
as seen in the drastic change of the electronic state with slight lattice deformation.
In fact, recent theoretical study suggests that the orbital order {\it follows} the structural transition at $T_{\rm MI}$ \cite{Gorelov10}.
This implies that the electronic state can be governed by a lattice correlation length scale, 
which is typically much longer than the electronic one.
As a consequence, the electronic system gains a long correlation length,
leading to the low-field nonlinear conduction phenomenon.

In summary, we report the nonlinear conduction property of the Mott insulating phase of Ca$_2$RuO$_4$.
To overcome the experimental difficulty associated with  heating,  
we utilized a non-contact thermometer to directly measure the sample temperature even in the presence of large Joule heating. 
The observed resistivity exhibits a clear nonlinearity characterized by a negative differential resistance at very low current range,
highly incompatible with conventional mechanisms of nonlinear conduction.
As an origin of the observed nonlinear conduction phenomena, 
we suggest an energy-gap suppression by current, as also seen in non-equilibrium superconducting or charge-order state.

We thank S. Ajisaka, S. Nakamura, Y. Nogami, T. Oka for stimulating discussion and 
M. Sakaki, Y. Kimura for experimental assistance.
This work was supported by a Grant-in-Aid for Scientific Research (B) and for Young Scientists (B) from JSPS,
and by a Grant-in-Aid for Scientific Research on Innovative Area ``Heavy Electrons'' from MEXT, Japan.


\begin{thebibliography}{99}

\item[$^\ast$] Email: okazaki.ryuji@cc.nagoya-u.ac.jp
\item[$^\dag$] Present address: Department of Physics, Meiji University, Kawasaki 214-8571, Japan

\bibitem{Imada98} M. Imada, A. Fujimori, and Y. Tokura: Rev. Mod. Phys. {\bf 70} (1998) 1039.
\bibitem{Oka03} T. Oka, R. Arita, and H. Aoki: Phys. Rev. Lett. {\bf 91} (2003) 066406.
\bibitem{Potember79} R. S. Potember, T. O. Poehler, and D. O. Cowan: Appl. Phys. Lett. {\bf 34} (1979) 405.
\bibitem{Tokura88} Y. Tokura, H. Okamoto, T. Koda, T. Mitani, and G. Saito: Phys. Rev. B {\bf 38} (1988) 2215. 
\bibitem{Asamitsu97} A. Asamitsu, Y. Tomioka, H. Kuwahara, and Y. Tokura: Nature (London) {\bf 388} (1997) 50.
\bibitem{Taguchi00} Y. Taguchi, T. Matsumoto, and Y. Tokura: Phys. Rev. B {\bf 62} (2000) 7015.
\bibitem{Zeng08} L. J. Zeng, H. X. Yang, Y. Zhang, H. F. Tian, C. Ma, Y. B. Qin, Y. G. Zhao, and J. Q. Li: Europhys. Lett. {\bf 84} (2008) 57011.
\bibitem{Tamura10} K. Tamura, T. Ozawa, Y. Bando, T. Kawamoto, and T. Mori: J. Appl. Phys. {\bf 107} (2010) 103716.
\bibitem{Guiot13} V. Guiot, L. Cario, E. Janod, B. Corraze, V. Ta Phuoc, M. Rozenberg, P. Stoliar, T. Cren, and D. Roditchev: Nat. Commun. 4:1722 (2013).

\bibitem{Nakatsuji97} S.~Nakatsuji, S. Ikeda, and Y.~Maeno: J. Phys. Soc. Jpn. {\bf 66} (1997) 1868.
\bibitem{Cao97} G. Cao, S. McCall, M. Shepard, J. E. Crow, and R. P. Guertin: Phys. Rev. B {\bf 56} (1997) R2916.
\bibitem{Alexander99} C. S. Alexander, G. Cao, V. Dobrosavljevic, S. McCall,  J. E. Crow, E. Lochner, and R. P. Guertin: Phys. Rev. B {\bf 60} (1999) R8422.

\bibitem{Nakatsuji00L} S.~Nakatsuji and Y.~Maeno: Phys. Rev. Lett. {\bf 84} (2000) 2666.
\bibitem{Nakatsuji00} S.~Nakatsuji and Y.~Maeno: Phys. Rev. B {\bf 62} (2000) 6458.
\bibitem{Nakamura02} F.~Nakamura, T. Goko, M. Ito, T. Fujita, S. Nakatsuji, H. Fukazawa, Y. Maeno, P. Alireza, D. Forsythe, and S. R. Julian:  Phys. Rev. B {\bf 65} (2002) 220402(R).

\bibitem{Mizokawa01} T. Mizokawa, L. H. Tjeng, G. A. Sawatzky, G. Ghiringhelli, O. Tjernberg, N. B. Brookes, H. Fukazawa, S. Nakatsuji, and Y. Maeno: Phys. Rev. Lett. {\bf 87} (2001) 077202.
\bibitem{Hotta01} T. Hotta and E. Dagotto: Phys. Rev. Lett. {\bf 88} (2001) 017201.

\bibitem{Anisimov02} V. I. Anisimov, I. A. Nekrasow, D. E. Kondakov, T. M. Rice, and M. Sigrist: Eur. Phys. J. B {\bf 25} (2002) 191.

\bibitem{Lee02} J. S. Lee, Y. S. Lee, T. W. Noh, S.-J. Oh, Yu. Jaejun, S. Nakatsuji, H. Fukazawa, and Y. Maeno: Phys. Rev. Lett. {\bf 89} (2002) 257402.
\bibitem{Jung03} J. H. Jung, Z. Fang, J. P. He, Y. Kaneko, Y. Okimoto, and Y. Tokura: Phys. Rev. Lett. {\bf 91} (2003) 056403.

\bibitem{Kubota05} M. Kubota, Y. Murakami, M. Mizumaki, H. Ohsumi, N. Ikeda, S. Nakatsuji, H. Fukazawa, and Y. Maeno: Phys. Rev. Lett. {\bf 95} (2005) 026401.

\bibitem{Zegkinoglou05}I. Zegkinoglou, J. Strempfer, C. S. Nelson, J. P. Hill, J. Chakhalian, C. Bernhard, J. C. Lang, G. Srajer, H. Fukazawa, S. Nakatsuji, Y. Maeno, and B. Keimer: Phys. Rev. Lett. {\bf 95} (2005) 136401.

\bibitem{Gorelov10} E.~Gorelov, M. Karolak, T. O. Wehling, F. Lechermann, A. I. Lichtenstein, and E. Pavarini: Phys. Rev. Lett. {\bf 104} (2010) 226401.

\bibitem{Maeno94} Y. Maeno, H. Hashimoto, K. Yoshida, S. Nishizaki, T. Fujita, J. G. Bednorz, and F. Lichtenberg: Nature (London) {\bf 372} (1994) 532.

\bibitem{Carlo12} J. P. Carlo, T. Goko, I. M. Gat-Malureanu, P. L. Russo, A. T. Savici, A. A. Aczel, G. J. MacDougall, J. A. Rodriguez, T. J. Williams, G. M. Luke, C. R. Wiebe, Y. Yoshida, S. Nakatsuji, Y. Maeno, T. Taniguchi, and Y. J. Uemura: Nat. Mater. {\bf 11} (2012) 323.

\bibitem{Alireza10} P. L. Alireza, F. Nakamura, S. K. Goh, Y. Maeno, S. Nakatsuji, Y. T. C. Ko, M. Sutherland, S. Julian, and G. G. Lonzarich: J. Phys.: Condens. Matter {\bf 22} (2010) 052202.


\bibitem{Nakamura11} F. Nakamura, M. Sakaki, Y. Yamanaka, S. Tamaru, T. Suzuki, and Y. Maeno: Sci. Rep. {\bf 3} (2013) 2536.

\bibitem{Ryder53} E. J. Ryder: Phys. Rev. {\bf 90} (1953) 766.
\bibitem{Glicksman58} M. Glicksman and M .C. Steele: Phys. Rev. {\bf 110} (1958) 1204.
\bibitem{Bauer72} G. Bauer and F. Kuchar: Phys. Stat. Sol. (a) {\bf 13} (1972) 169.

\bibitem{Sacanell05} J. Sacanell, A. G. Leyva, and P. Levy: J. Appl. Phys. {\bf 98} (2005) 113708.
\bibitem{Song06} H. Song, M. Tokunaga, S. Imamori, Y. Tokunaga, and T. Tamegai: Phys. Rev. B {\bf 74} (2006) 052404.
\bibitem{Carneiro06} A. S. Carneiro, R. F. Jardim, and F. C. Fonseca: Phys. Rev. B {\bf 73} (2006) 012410.
\bibitem{Hucker07} M. H\"ucker, M. v. Zimmermann, and G. D. Gu: Phys. Rev. B {\bf 75} (2007) 041103(R).
\bibitem{Kim10} J. Kim, C. Ko, A. Frenzel, S. Ramanathan, and J. E. Hoffman: Appl. Phys. Lett. {\bf 96} (2010) 213106.
\bibitem{Zhong11} X. Zhong, X. Zhang, A. Gupta, and P. LeClair: J. Appl. Phys. {\bf 110} (2011) 084516.
\bibitem{Zimmers13} A. Zimmers, L. Aigouy, M. Mortier, A. Sharoni, S. Wang, K. G. West, J. G. Ramirez, and I. K. Schuller: Phys. Rev. Lett. {\bf 110} (2013) 056601.

\bibitem{Puchkov98} A. V. Puchkov, M. C. Schabel, D. N. Basov, T. Startseva, G. Cao, T. Timusk, and Z.-X. Shen: Phys. Rev. Lett. {\bf 81} (1998) 2747.
\bibitem{Nakatsuji04} S. Nakatsuji, V. Dobrosavljevi\ifmmode \acute{c}\else \'{c}\fi{}, D. Tanaskovi\ifmmode \acute{c}\else \'{c}\fi{}, M, Minakata, H. Fukazawa, and Y. Maeno: Phys. Rev. Lett. {\bf 93} (2004) 146401.

\bibitem{Braden98} M. Braden, G.  Andr\'e, S. Nakatsuji,  and Y. Maeno: Phys. Rev. B {\bf 58} (1998) 847.

\bibitem{Fridet01} O. Friedt, M. Braden, G. Andr\'e, P. Adelmann, S. Nakatsuji,  and Y. Maeno: Phys. Rev. B {\bf 63} (2001) 174432.


\bibitem{Oka12} T. Oka: Phys. Rev. B {\bf 86} (2012) 075148.

\bibitem{Conwellreview} E. M. Conwell: Solid State Phys. Suppl. {\bf 9} (1967) 81.

\bibitem{Wolff54} P. A. Wolff:  Phys. Rev. {\bf 95} (1954) 1415.
\bibitem{Owen72} C. S. Owen and D. J. Scalapino: Phys. Rev. Lett. {\bf 28} (1972) 1559.

\bibitem{Fuchs77} J. Fuchs, P. W. Epperlein, M. Welte, and W. Eisenmenger: Phys. Rev. Lett. {\bf 38} (1977) 919.
\bibitem{Akoh82} H. Akoh and K. Kajimura: Phys. Rev. B {\bf 25} (1982) 4467.

\bibitem{Ajisaka09} S. Ajisaka, H. Nishimura, S. Tasaki, and I. Terasaki: Prog. Thoer. Phys. {\bf 121} (2009) 1289.
\bibitem{Sawano05} F. Sawano, I. Terasaki, H. Mori, T. Mori, M. Watanabe, N. Ikeda, Y. Nogami, and Y. Noda: Nature (London) {\bf 437} (2005) 522.


\end{thebibliography}
\end{document}